**Glutathione conjugates of the mercapturic acid pathway and guanine adduct as biomarkers of exposure to CEES, a sulfur mustard analog.**

By Marie Roser[1], David Béal[1], Camille Eldin[1], Leslie Gudimard[1], Fanny Caffin[2], Fanny Gros-Désormeaux[2], Daniel Léonço[3], François Fenaille[3], Christophe Junot[3], Christophe Piérard[2] and Thierry Douki[1]*

1) Univ. Grenoble Alpes, CEA, CNRS, IRIG, SyMMES, F-38000 Grenoble
2) Institut de Recherche Biomédicale des Armées (IRBA), Place Général Valérie André, BP 73, 91223 Brétigny-sur-Orge Cedex, France
3) Université Paris-Saclay, CEA, INRAE, Département Médicaments et Technologies pour la Santé (DMTS), MetaboHUB, F-91191 Gif sur Yvette, France

* Corresponding author:

Thierry Douki

ORCID: 0000-0002-5022-071X

Email : thierry.douki@cea.fr

Tel (33) 4 38 78 31 91

Fax (33) 4 38 78 50 90

**Abstract**

Sulfur mustard (SM), a chemical warfare agent, is a strong alkylating compound that readily reacts with numerous biomolecules. The goal of the present work was to define and validate new biomarkers of exposure to SM that could be easily accessible in urine or plasma. Because investigations using SM are prohibited by the Organization for the Prohibition of Chemical Weapons, we worked with 2-chloroethyl ethyl sulfide (CEES), a monofunctional analog of SM. We developed an ultra-high-pressure liquid chromatography - tandem mass spectrometry approach (UHPLC-MS/MS) to the conjugate of CEES to glutathione and two of its metabolites, the cysteine and the N-acetyl-cysteine conjugates. The N7-guanine adduct of CEES (N7Gua-CEES) was also targeted. After synthesizing the specific biomarkers, a solid phase extraction protocol and a UHPLC-MS/MS method with isotopic dilution were optimized. We were able to quantify N7Gua-CEES in the DNA of HaCaT keratinocytes and of explants of human skin exposed to CEES. N7Gua-CEES was also detected in the culture medium of these two models, together with the glutathione and the cysteine conjugates. In contrast, the N-acetyl-cysteine conjugate was not detected. The method was then applied to plasma from mice cutaneously exposed to CEES. All four markers could be detected. Our present results thus validate both the analytical technique and the biological relevance of new, easily quantifiable biomarkers of exposure to CEES. Because CEES behaves very similarly to SM, the results are promising for application to this toxic of interest.

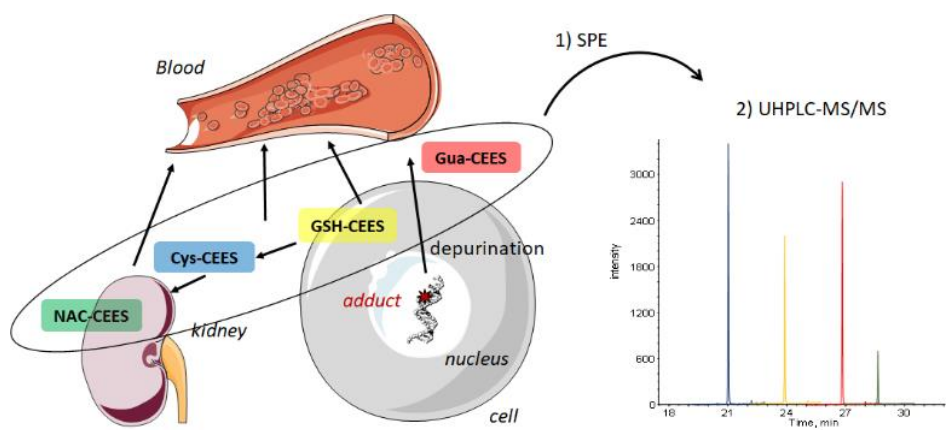

# Introduction

Sulfur mustard (SM) was first used as a chemical warfare agent during World War I [1] and on other battlefields since then. SM was even used in recent conflicts in the Middle East, although the Organization for the Prohibition of Chemical Weapons (OPCW) prohibits the synthesis of chemical warfare agents and restricts their use to protection and peaceful research. In addition to these military concerns, SM is also a threat for civilians. Many SM-containing ammunition may be found buried or abandoned at former defense sites. Large amounts of SM have also been disposed at the bottom of the sea. More importantly, SM is thought to be a persistent threat by non-state actors, such as terrorist groups, because of its easy synthesis and handling, and the risk of theft from stockpiles [2].

The severity of SM-induced complications mainly depends on dose, duration, and body part of the exposure [3, 4]. SM contamination may cause adverse health effects in many organs especially the skin, eyes, and respiratory system [3, 4]. The first symptoms, including ocular, dermal, and respiratory damage, usually become evident only approximately 12 hours after exposure [5]. SM is classified as a vesicant because it causes blistering of the skin as a major feature of its toxic effects observed shortly after exposure. SM also exhibits long-term effects. Local effects of SM exposure may include scarring of the skin and bronchial stenosis [6]. Other long-term effects may appear some months or years after the exposure and have a serious impact on the life quality of the exposed individuals [7]. SM is also a known carcinogen [8, 9]. Unfortunately, there is no antidote available for preventing the effects of SM and no specific therapy for the treatment of SM-induced lesions. The aim of current therapeutic strategies is to relieve symptoms, prevent infections, and promote healing.

Some cellular and molecular processes have been identified that partially explain the toxicological effects of SM, but the overall exact mechanism remains unclear [10]. It should be stressed that because synthesis, use and storage of SM are strongly regulated by the OPCW, a large fraction of the available data have been obtained with the analog of SM (2-chloroethly)-ethyl-sulfide (CEES), a monofunctional molecule bearing one chlorine atom instead of two. SM and CEES induce similar severe burns on skin in animals. They are both strong alkylating agents, mostly because they undergo intramolecular cyclization into episulfonium ions. These electrophilic species can readily react with water and nucleophilic sites present in cellular macromolecules such as glutathione, DNA and proteins (Fig. 1). Most products of these reactions are stable adducts or conjugates which can affect the normal functions of macromolecules, such as the enzymes involved in cell energetic metabolism. Some metabolites are excreted in urine, feces and blood [11, 12]. These SM-specific derivatives can be easily accessible targets in biological fluids and used as specific biomarkers. Actually, detection of such analytes in urine or blood is necessary for many reasons such as in forensic science for providing evidence of contamination, or in therapeutics for diagnostic and prognostic purposes.

< Figure 1>

A first proposed approach for biomonitoring of exposure to SM involved quantification of the direct oxidation product of SM, bis-chloroethyl sulfoxide (SMO). Rapid, simple and quantitative analytical methods were developed for this early SM biomarker present in blood for a rapid diagnostic during the asymptomatic latency period (less than 12h) [13, 14]. The directly hydrolyzed metabolite thiodiglycol (TDG) was detected in urine from SM-treated rats for 48h [15]. Under its bound and its oxidized forms, it was present up to 7 day in the blood and urine of the animals. In another study, TDG and TDGO were detected in the plasma of SM exposed rats, and the time windows for quantitative detection were 4 and 12h respectively [16]. Interestingly, similar results have been obtained in human patients [17]. Much debate has concerned the validity of TDG and TDGO as biomarkers because both can be detected at trace levels in unexposed human populations [11].

The second type of SM biomarkers involved products of its reaction with various amino acid residues present in proteins, including the 2-(2-hydroxyethyl)-thioethyl-valine adduct of hemoglobin and 2-(2-hydroxyethyl)-thioethyl-cysteine adduct of albumin. They can be detected in appropriate samples such as blood, serum, plasma or urine [17, 18]. In recent years, studies were more focused on albumin adducts [18–20]. The time-course studies in rats showed that the hemoglobin adduct is far more persistent than the albumin adduct. A study showed that the adduct to the N-terminal valine in hemoglobin was still present 28 days after exposure [21]. This long persistence makes this conjugate relevant for a retrospective detection. However, the sample preparation requires a time-consuming digestion step for the selective cleavage of the adduct from the protein.

Another major pathway in the toxicity of SM is its reaction on the critical nucleophilic sites of DNA to produce SM-DNA adducts (Fig. 1). To date, four major types of DNA adducts have been reported. These include the mono-alkylated base adducts 2-(2-hydroxyethyl)-thioethyl-N7-guanine (N7Gua-HETE), 2-(2-hydroxyethyl)-thioethyl-N3-adenine (N3Ade-HETE), the SM-guanine-guanine biadduct and an adduct involving both glutathione and guanine [22–25]. Formation of these adducts was extensively studied within DNA extracted from skin and internal organs after exposure [26–30]. DNA adducts appear to be valuable biomarkers since they are persistent, especially the N7Gua derivative which is the most abundant and still detected 3 weeks after exposure of hairless mice [27]. This was confirmed by the unambiguous detection of adducts in the DNA extracted from blood cells form exposed patients [17]. DNA adducts were also detected for several days in urine under the form of modified bases following spontaneous depurination or release by DNA repair systems [17, 31].

A last class of reaction products of biomolecules with SM that can be used as biomarkers results from the conjugation of SM to glutathione (GSH), an abundant cellular tripeptide. GSH plays a major role in the chemical inactivation of reactive electrophilic toxicants or metabolites either spontaneously or by catalysis by glutathione-*S*-transferases in the liver and in other organs such as the skin [32, 33]. The product is a glutathione-*S*-SM -conjugate (GSH-SM). In the complex pathway of glutathione metabolism [34], GSH-SM is consecutively converted into the cysteinyl-glycine-*S*-conjugate, the cysteine-*S*-conjugate and N-acetylcysteine-*S*-conjugate (NAC-SM) (Fig. 1). This last metabolite, also known as mercapturate derivative, is generally more polar and water soluble than the unconjugated electrophiles. It is thus readily excreted and eliminated in urine and/or bile [35, 36]. In terms of GSH-derived biomarkers, available data mostly involved the detection of glutathione-SM metabolites arising from β-lyase cleavage, leading to the formation of SBSNAE, SBMTE, SBMSE [17, 37, 38]. These metabolites are only detected for a few days in patients, and not in all of them [17, 39]. Surprisingly, other metabolites of the mercapturic acid pathway such as NAC-SM have not been used as biomarkers while they are commonly detected in biological fluids following exposure to other toxic compounds [40–43]. Data have been reported that show a larger amount of Cys conjugates in the bile than derivatives of the β-lyase pathway [44], strongly suggesting the formation of other GSH-SM metabolites such as the NAC conjugate.

This brief overview of the available biomarkers of exposure to SM shows that none is perfect, in terms of either specificity, persistence or analytical throughput. Moreover, proof of exposure is more reliable with a combination of information based on different biomarkers. There is thus room for additional biomarkers. To this end, we decided to explore the possibility of using the conjugates of the mercapturate pathway with CEES as a SM analog. As explained above, CEES was chosen because use of SM is strictly limited. We felt necessary to first develop our approach with an analog and then, based on this first work, rapidly extend the method to SM. Our aim was to quantify conjugates between CEES and GSH, cysteine and N-acetylcysteine (GSH-CEES, Cys-CEES and NAC-CEES, respectively) (Fig. 2). The chemical structure of these compounds have been established for SM in previous works [45, 46]. We also included the DNA adduct (2-ethyl)-thioethyl-N7Gua (N7Gua-CEES) to our selected biomarkers since promising results have been obtained with the similar SM-derivative. We first synthesized the standards and their isotopically labelled analogs, and developed an ultra-high-pressure liquid chromatography - tandem mass spectrometry approach (UHPLC-MS/MS) with isotopic dilution. Then, the clean-up step of the samples was optimized. Finally, we evaluated the biological relevance of these biomarkers in cultured cells, human skin explants and blood plasma from mice.

< Figure 2 >

## Materials and methods

**Chemicals and enzymes**

Reduced L-glutathione (GSH), (2-chloroethyl)-ethyl sulfide (CEES) and L-cysteine (Cys) were purchased from Sigma Aldrich (Saint Quentin Falavier, France). N-Acetyl-L-cysteine (NAC) was purchased from Roche (Boehringen Mannheim) (Mannheim, Germany). 2'-Deoxyguanosine (dGuo) was purchased from Pharma Waldhof (Düsseldorf, Germany). Formic acid (LC-MS grade) was purchased from Thermo Fisher Scientific (Rockford, USA), acetonitrile (HPLC-MS grade) from VWR (Fontenay-sous-Bois, France), methanol (HPLC grade) and ammonium formate from Sigma Aldrich (Saint Quentin Falavier, France). Ribonuclease T1, ribonuclease A, phosphodiesterase II, deoxyribonuclease II, alkaline phosphatase and nuclease P1 were obtained from Sigma. Protease was purchased from Qiagen and phosphodiesterase I from Worthington (Lakewood, NJ, USA). Isotopically labelled molecules Cys* ($^{13}C_3$ 99%, $^{15}N$ 99%), NAC* ($^{13}C_3$ 97-99% $^{15}N$ 97-99%), GSH* ($^{13}C_2$ 98% $^{15}N$ 96-99%) and dGuo* ($^{15}N_5$ 98%) were purchased from Cambridge Isotope Laboratories (Andover, USA).

**Synthesis of biomarkers**

Aqueous solutions of either 10.8 mg of Cys, 10.9 mg of NAC, 10 mg of dGuo or 10.4mg of GSH were prepared in 5mL of phosphate buffer (10 mM, pH=7). Five µL of CEES (2.9 mM) was added to each tube. The reaction lasted overnight. Several washing steps with dichloromethane were carried out to eliminate possible traces of CEES. Depuration of N7Gua-CEES adducts was completed by thermal hydrolysis (90°C for 20 min in a heating block) at neutral pH as previously reported [47]. All biomarkers were isolated by using a preparative HPLC system consisting of a L7100 pump from Merck-Hitachi, a 7125 injector valve from Rheodyne (San Jose, USA), a C18 reverse phase column (4.0x250mm, 5 µm particle size, Uptisphere, Interchim, Montluçon, France), a L4200 UV-Vis detector from Merck (Darmstadt, Germany). A gradient of ammonium formate (AmF, 5mM) and acetonitrile (0 min: 0% ACN, 10 min: 5% ACN, 30 min: 20% ACN, 45 min: 30% ACN) at a flow rate of 1mL/min was used. Fractions of 2 mL were collected from which 20 µL was analyzed by UHPLC-MS/MS. The fractions of interest were pooled and further purified on the HPLC system until a pure product, as inferred from HPLC profiles recorded at 220 nm, was obtained. The final solutions were freeze-dried and suspended in MilliQ water. N7Gua-CEES adduct was identified by comparison with a previously synthesized product [47]. GSH, NAC and Cys conjugates were also characterized by $^1$H-NMR in D$_2$O. The concentration of each standard was determined by UHPLC-MS[1] used in the single ion monitoring. For each conjugate, the reference compound was the parent molecule expected to exhibit the same ionization efficiency. Results were confirmed by analysis in UHPLC with diode array detector. For N7Gua, calibration was also performed by UV absorption of the standard solution using the molar absorption coefficient of N7-methyl guanine for the calculation.

**Synthesis of internal standards (IS)**

Solutions of 10 mM of each labelled biomolecules (either Cys *, NAC*, GSH* or dGuo*) were prepared in 1/10 PBS. CEES (20 µL) was added to 5 mg of the labelled compounds, namely either 4 mL of Cys*, 1.61 mL of GSH*, 1.72 mL of dGuo* or 3 mL NAC*. The reaction lasted overnight. Several washing steps with dichloromethane were carried out to eliminate possible traces of CEES. All biomarkers were isolated by using SPE (see below). One mL fractions containing increasing proportion of methanol were collected and 20 µL of each fraction was analyzed by UHPLC-MS/MS. The fractions of interest were pooled. The final solutions were freeze-dried and suspended in MilliQ water. The concentration of each internal standard was determined by UHPLC-MS/MS calibrated with unlabeled compounds.

**Sample preparation: Solid Phase Extraction (SPE) of biomarkers from culture medium and plasma**

Prior to SPE purification, all samples were centrifuged at 5000xg and the liquid phase was collected. Sample preparation, *i.e.* clean-up and concentration, was performed on Chromabond® HR-X 100 (45 µm particle size, 100 mg polymeric phase) SPE columns using a vacuum manifold. Columns were conditioned with 1 mL of methanol containing 0.1% formic acid (FAc), followed by elution of 0.1% FAc in water (1 mL). The samples to be analyzed (500 µL of culture medium or 100 µl of plasma) were spiked with internal standards (10 µL, 0.2 mM) and loaded onto the SPE column. The sample was washed with 0.1% formic acid (FAc) in water (1 mL) followed by 5% methanol containing 0.1% FAc (1mL). All traces of washing solution was removed from the column by suction for 10 minutes under vacuum. In the final step, the analytes of interest were eluted into a hemolysis tube with 1 ml of a buffer prepared by dilution of 200 µL of a 10 M ammonium formate (AmF) into 200 mL pure methanol. The eluent was dried in a Speed Vac at 35°C. The samples were reconstituted in water (200 µL) and stored at -20°C until UHPLC-MS/MS analysis.

**UHPLC separations**

Chromatographic separations were carried out on a Sciex ExionLC UHPLC system (Framingham, MA) equipped with an autosampler, a binary pump, and a heated column compartment. Analytical separation was performed using a Macherey-Nagel Nucleodur® C18 column (100 mm x 2.0 mm ID, 1.8 µm particle size). Mobile phase A consisted of 2mM AmF and 0.2% FAc in MilliQ water and mobile phase B was LC/MS grade acetonitrile.

For analysis of N7Gua-CEES in DNA, the initial conditions of the gradient were 2% B for 5 minutes with the flow rate was set to 350 µL/min. The column temperature maintained at 40°C and the sample storage compartment set to 15°C. The proportion of acetonitrile increased to 8% between 5 and 8 min, followed by a deeper linear increase to 40% between 8 and 11.5 min, an immediate ramp to 95% for 1 min followed by a return back to initial conditions at 12.6 min. An equilibration phase was maintained until 15 min. The amount of normal DNA nucleosides was quantified by a UV detector. Quantification of DNA lesions and normal nucleosides were performed by external calibration using calibrated solutions of authentic standards. Results were expressed as the number of adduct per million normal bases.

For the quantification of biomarkers in the culture medium, the initial conditions of the gradient were 5% B for 1.2 minutes. The flow rate was set to 400 µL/min, the column temperature maintained at 50°C, and the temperature of the sample storage compartment set to 15°C. A linear increase of the solvent to 35% was applied between 1.2 and 7.2 min, followed by a ramp to 70% B for 1 min and ten by a return back to initial conditions at 8.4 until 11 min. Quantitative results were obtained using internal standards. Indeed, the response of the analyte is normalized to the response of the IS. The ratio of the peak area of the analyte in the sample to the peak area of the IS in the sample is compared to a similar ratio derived for each calibration standard.

**Mass spectrometry detection**

The UHPLC system was coupled to a SCIEX QTRAP® 6500+ triple quadrupolar mass spectrometer (Framingham, MA) equipped with an Ion Drive™ Turbo V source operated in positive ionization mode. Data were collected in multiple reaction monitoring (MRM) scan mode. Transitions corresponding to the three most intense fragments were used as confirmation ions and the most sensitive one among those was used for quantification. Three MRM transitions were recorded for each of the isotopically labelled compounds corresponding to the transitions of their native analogs (Electronic Supplementary Material, Table S1). The following optimized ionization parameters were applied for all analytes: curtain gas (45 psi), collision gas (70 psi), IonSpray voltage (5300 V), temperature (550°C), Ion Source Gas 1 (60 psi), Ion Source Gas 2 (60 psi). Other compound-dependent parameters are listed in table S1 below. Data acquisition and quantitative analyses were performed using the Analyst and MultiQuant softwares, respectively.

< Electronic Supplementary Material, Table S1 >

**Treatment of HaCaT human keratinocyte cell line**

HaCat human keratinocytes (ThermoFicher Scientific) were grown in DMEM-GlutaMAX™ (Gibco) medium supplemented with 10% (v/v) fetal calf serum (Gibco), and 1% (v/v) of a mixture of antibiotics (50 U/mL of penicillin and 50 mg/mL of streptomycin). Cells, seeded at 14 300 / cm$^2$ in 60 mm diameter petri dishes, were grown in a Hera Cell incubator (Heraeus) in a 5% $CO_2$ atmosphere at 37°C. Treatments were performed in triplicate for each CEES concentration (0, 1, 2, 5, and 10 mM). Before treatment, the medium was discarded. Then, cells were suspended for 30 min at room temperature in 3 mL DPBS (Gibco) containing the different CEES concentrations. Subsequently, they were washed twice with PBS to remove traces of CEES. Cells were thereafter incubated for 6h at 37°C with "fresh" medium. Last, the medium was collected and stored at -20°C before purification by solid phase extraction. The cells were collected by trypsination. After deactivation of trypsin by addition of medium, the cells were centrifuged (1500 rpm, 5 min) in a Megafuge 1.0R centrifuge (Heraeus) and the supernatant was discarded. The pellet was suspended in 1 mL PBS and placed in a 2 mL Eppendorf tube. Samples were centrifuged (400xg, 5min) and the supernatant was discarded. The HaCaT pellets were stored at -20°C before DNA extraction.

DNA extraction from CEES-treated HaCaT was carried out following the NaI/propan-2-ol chaotropic method [48]. Briefly, plasmic membranes were lysed by adding a buffer (sucrose 320 mM, $MgCl_2$ 5 mM, Tris-HCl 10 mM, deferoxamine 0.1 mM) containing 1% (v/v) triton X-100. The nuclei were collected by centrifugation in a Megafuge 40R centrifuge (Heraeus) (5 min at 1500xg). Then, the nuclear membranes were lysed by adding a second buffer (EDTA-$Na_2$ 5 mM, Tris-HCl 10 mM, deferoxamine 0.15 mM) containing 0.5% (v/v) SDS. RNA was hydrolyzed using RNases A (1.5 µL, 100 mg/mL) and T1 (3.5 µl, 1 U/µl) at 37 °C for 15 min. Samples were then incubated at 37°C for 1 h after addition of protease. DNA was precipitated with 2 volumes of buffered NaI-solution (EDTA-$Na_2$ 20 mM, NaI 7.6 M, Tris-HCl 40 mM, deferoxamine 0.3 mM) and 3.3 volumes of absolute propan-2-ol. The DNA pellet was then washed by successive centrifugation and resuspension, using first aqueous propan-2-ol (40%, v/v) and then aqueous ethanol (70%, v/v).

DNA was hydrolyzed as previously described by Douki et al [49] in two steps. Each involved a specific enzymatic cocktail: nuclease P1, phosphodiesterase II and DNase II at pH 6 (50 mM succinic acid, 25 mM $CaCl_2$, 75 mM ammonium acetate) for the first; phosphodiesterase I and alkaline phosphatase at pH 8 (50 mM tris) for the second. Each step was carried out at 37°C for 2 h. The samples were then left at 90°C for 20 min in a heating block in order to depurinate DNA adducts. The final samples thus contain normal DNA bases under the form of nucleosides, which are quantified by a UV detector placed before the inlet of the mass spectrometer, and adducts as modified bases quantified by mass spectrometry.

**Treatment of skin explants**

Human skins were obtained immediately after breast plastic surgery from healthy donors with their informed consent (Centre Hospitalier Universitaire de Grenoble, France), in agreement with article L1245-2 of the French Public Health Code on the use of surgical wastes for research purposes. Declaration of collection and storage of human skin was recorded in the CODECOH DC-2019-3391 document, which was validated by the French ministry of Research (June 17, 2019). All donors were Caucasian (16–62 year old), and their skin phototype was either II or III according to the Fitzpatrick classification. After surgery, the skin was immediately transported to the laboratory at ambient temperature in 50 mL falcon tubes (Becton Dickinson Labware, USA). Skin was disinfected with PBS supplemented with 0.4% betadine for 15 min and rinsed twice with PBS containing 10% Pen/Strep. The skin, from which the fat layer was removed but that still contained both the dermis and the epidermis, was cut into 12-mm diameter discs with sterile punches (Help Medical, France). Skin biopsies were placed with the dermal side down into ThinCert™ inserts (14mm inner diameter, 1µm pore size filter, Greiner Bio-One, Austria) maintained in 12 well cell culture plates (sterile, with lid) (Greiner Bio-One, Austria). Medium (600 µL) was added into the wells, under the inserts. The culture medium used was DMEM-F12 supplement with 1% Pen/Strep. Four doses of CEES (0, 2, 4, 10 mg/ml) in solution in 5µL of dichloromethane were applied to the epidermal surface, corresponding to 0, 0.08, 0.16, 0.40 µmol, respectively. Dichloromethane evaporation was permitted by waiting 30s and then the lid was replaced. Afterwards, skin biopsies were placed at 37°C in 5% $CO_2$ air in the incubator for three different time points: 1, 6 and 24h. Collected medium from samples was directly stored at -20°C until SPE. Biopsies were also frozen at -20°C until DNA extraction. Experiments were performed in triplicates and were repeated on skins from two different donors.

DNA was extracted by using the DNEasy Tissue Kit obtained from Qiagen (Courtaboeuf, France). Briefly, the first step was the grinding of the biopsy in buffer ATL (350µL) by using a TissueLyzer (15min, 25Hz). After adding proteinase K, samples were incubated 3h at 55°C. An RNase A treatment and a second lysis step involving buffer AL were performed (10 min at 70°C) for a complete lysis of the tissue. Lysate samples were then loaded onto DNEasy mini spin columns. After two washing steps, DNA was eluted with 2x100 µl of deionized water. The sample was freeze-dried overnight, and the resulting DNA residue was dissolved in 50 µL of water. DNA was hydrolyzed as described for DNA extracted from HaCat cells.

**Animal experiment**

The protocol was validated by the ethic committee of Health Services of French Armed Forces, according to the French 2013-118 decree that follows the 2010/63 directive from the EU on animal research. SKH-1 hairless mice (n=5) were purchased from Charles River and used at 7 weeks. Mice were anesthetized by intraperitoneal injection of a mixture of ketamine (90 mg/kg) and xylazine (10 mg/kg). Under a fume hood, mice were then locally exposed through the skin to the vapors emitted by 10 µL of neat CEES using cap method (screw cap for HPLC vial, PTFE septum and polypropylene cap). Four sites on the mice back were exposed to CEES vapors (Electronic Supplementary Material, Fig. S1). The caps containing CEES were removed after 30 min. After 4h, the skin was decontaminated by soap and water. To relieve pain induced by CEES, buprenorphine (0.05 mg/kg) was subcutaneously injected after skin decontamination and codeine was daily administrated in drinking water (4 mg/100 mL). One day after the exposure, the animal were sacrificed by an injection of pentobarbital (150 mg/kg, ip route) and the blood was collected by heart puncture. Plasma was separated by centrifugation (2000 g, 10 minutes, 4°C) and samples were kept frozen until quantification of biomarkers. For this purpose, an aliquot fraction of 100 µl was collected from each sample. The SPE and UHPLC-MS/MS conditions were the same as those applied to cell culture and explants medium.

< Electronic Supplementary Material, Figure S1 >

## Results and discussion

**Method optimization**

*Characterization of the standard molecules.*

The standards used for the method optimization and the quantitative analysis of the samples were synthetized by incubation of the parent compound with CEES followed by HPLC purification. The material present in the collected fractions was then characterized. $^1$H-NMR spectra were recorded for the GSH, Cys and NAC conjugates. They exhibited all the expected signals for the analytes (Electronic Supplementary Material, Table S2), thereby providing evidence that all expected compounds had been successfully synthesized. Signals corresponding to traces of impurities were detected for Cys-CEES and NAC-CEES. However, the purity of these standards was satisfactory, reaching 95 and 83 % respectively. $^1$H-NMR analysis showed that GSH-CEES was the major product in the standard solution but that other minor compounds were present. The solution was thus analyzed in HPLC-MS with recording of the full MS1 spectrum over the 200-450 m/z range. Compounds exhibiting a pseudo molecular ion at m/z = 371, 415, 245 and 261 were observed on addition to GSH-CEES (m/z 396). The latter compounds was the most concentrated, in agreement with the NMR data. The concentration of second next product was half of that of GSH-CEES, while the proportion of the other impurities was below 10%. Since the mass of the contaminants was different from that of GSH-CEES, no interference was possible the MS$^1$ single ion monitoring used for the calibration of the solutions. No impact was expected either in the MRM detection mode used for the samples, as confirmed by the observation that GSH-CEES was the only compound detected in the product ion scan mode ($MS^2$ fragmentation). We did not attempt to purify further the standard of GSH-CEES in order to prevent loss of material. The amount of N7Gua-CEES was too low to allow recording of a NMR spectrum. We are yet confident in the identity of the analyte since all chromatographic, mass spectrometry and UV absorption features are identical to those for a previous synthesis validated in reference [47]. In particular, the UV absorption spectrum exhibited the typical blue shifted maximum of N7-alkylated derivative of guanine. This absorption was used to calibrate the obtained solution.

< Electronic Supplementary Material, Table 2 >

*UHPLC-MS/MS quantification*

Availability of authentic pure standard in solution allowed us to optimize the parameter of the liquid chromatography-mass spectrometry detection. Standard Liquid chromatography conditions were chosen to insure a proper elution of the analytes. UHPLC was favored over HPLC in order to reduce the analysis time and increase the throughput of the method. Standard chromatographic parameters were used, in terms of both column type and gradient. Detection was afforded by a triple quadrupolar mass spectrometer used with positive ionization and in the "reaction monitoring mode". This approach relies on the monitoring of specific fragmentation reactions of the pseudo-molecular ion of targeted analytes. The observed fragmentations (Electronic Supplementary Material, Fig. S2) were mostly observed on the CEES moiety of the analytes with formation of fragments both at m/z=61 ($C_2H_5S$) and 89 ($C_4H_9S$) for all compounds. One exception was GSH-CEES for which the former fragment was replaced for by a more intense one at m/z=121 ($C_4H_9S_2$). Fragments at m/z=321 (loss of $CO_2H-CH-NH_2$) was observed for GSH-CEES and at m/z=120 (loss of $C_4H_9S$) for Cys-CEES. Detection in the MRM mode is very specific and, because it leads to a low background noise, very sensitive.

< Electronic Supplementary Material, Figure S2 >

*Sample preparation*

Solid phase extraction (SPE) is routinely used in sample preparation for quantifying analytes in biofluids such as plasma and urine. This technique allows both removal of interfering biological matrix components and increase of the concentrations of the analytes in the final samples. Application of this method to organic compounds in solution in water or biofluids requires adsorption on proper solid material followed by desorption with a small quantity of an organic solvent. We tried different sorbents, namely C18 and polymeric phases. The polymeric phase afforded the best recovery with less variability. We also optimized the composition of the liquid phases, mostly the impact of FAc on the different steps of the sample preparation. This acid improves the retention of the analytes on the sorbent. The optimized protocol is described in the Materials and Methods part. For validation purposes, 5 samples of matrix blank (HaCaT culture medium) were spiked with IS and standards (5 pmol) to obtain a concentration in the middle of the calibration curve typically used for the biological experiments (from 0.1 to 50 pmol). The yields of the complete sample preparation *i.e.* from the collection to the analysis include SPE, drying step, sample reconstitution were 65±5%, 84±3% 112±4% 106±5% for GSH-CEES, Cys-CEES, NAC-CEES and N7Gua-CEES, respectively.

It should be emphasized that the stability of the targeted analytes was a major concern during the method's optimization. For example, formic acid was removed from the SPE elution solvent because it induced some degradation during the evaporation. We also checked the long-term stability of the analytes in the matrices. They were found to be stable in culture medium at -20°C for at least 6 months, with the exception of Gua-CEES that underwent 50% degradation over this period (data not shown). Last, we investigated whether oxidized derivatives of the analytes could be produced. Previous reports of the detection of oxidation products of SM [13, 14] suggested that it could also take place for CEES biomarkers. Therefore, we analyzed cultured medium from CEES-treated skin explants (*vide infra*) in the product ion scan mode. This approach provides fragmentation mass spectra and thereby structural information on components of a mixture. Emphasis was placed on the Cysteine conjugate, which was the biomarker detected in the largest concentration in this experiment. An intense peak was detected for the fragmentation of the pseudo-molecular ion of Cys-CEES. In contrast, none was observed for masses with one (M+16) or 2 (M+32) oxygen atoms that would correspond to sulfoxides (-SO-) and sulfones (–$SO_{2-}$), respectively (Electronic Supplementary Material, Fig. S3).

< Electronic Supplementary Material, Figure S3 >

*Method validation*

Validation of the method was performed by following the FDA recommendations for Bioanalytical Method Validation [50] and the Standard Practices for Method Validation in Forensic Toxicology [51]. The limit of detection (LOD) and limit of quantification (LOQ) for each biomarker were established by using the method called "blank determination" according to *Shrivastana et al.* [52]. The blank determination is applied when the blank analysis gives results with a nonzero standard deviation. LOD is expressed as the analyte concentration corresponding to the sample blank value plus three standard deviation and LOQ is the analyte concentration corresponding to the sample blank value spiked with IS plus ten standard deviations. The values were determined with 18 matrix blank samples (cell culture medium). The calculated LOD for each analyte was 9, 48, 46 and 37 fmol for GSH-CEES, Cys-CEES, NAC-CEES and N7Gua-CEES, respectively. The corresponding calculated LOQ were 22, 111, 120 and 88 fmol.

This calculation method was found to be reliable since the determined LOD matches with the lowest concentration leading to an observable peak on the chromatogram. In order to define a proper calibration curve, the lowest concentration we used for the standards was 100 fmol, namely in the range of mean LOQ.47

Accuracy and precision were evaluated by analyzing intraday and interday variations for blank, zero calibrator, and six calibrators covering the quantitation range including LOQ (Electronic Supplementary Material, Table S3). The matrix was culture medium from unexposed HaCaT cells. The concentration-response relationship for each analyte was fitted by a simple linear regression forced through zero. Our results are in accordance with the FDA criteria because non-zero calibrators are ± 15% of nominal (theoretical) concentrations, and for the lowest concentration ± 20% of the nominal concentrations in each validation run. The selectivity was also studied. Blank and zero calibrators were free of interference at the retention times of the analytes and the IS. The method was thus found to be highly specific, mostly as the result of the different determination criteria by MRM (ion quantitation and ion ratio). No carryover between samples was observed, even with the highest calibrator. Limited amount of sample from untreated mice prevented us from performing a complete validation in plasma. Yet we were able to show the good intraday reproducibility (with n=4) on a mixture of several samples from exposed animals (Electronic Supplementary Material, Table S4).

< Electronic Supplementary Material, Table S3 >

< Electronic Supplementary Material, Table S4 >

**CEES biomarkers in cultured HaCaT cells**

Skin is a direct target organ of SM. Therefore, we performed a first *in vitro* biological experiment with a human keratinocyte cell line (HaCaT). Indeed, keratinocyte represents the major cell type in epidermis, the upper layer of skin. Both adduct in nuclear DNA and conjugates in the medium were targeted. It was not possible to treat cells under growing conditions because the medium contains amino acids that would lead to false positive detection. Therefore, rather than treating cells for a long time in their culture medium with micromolar concentrations of CEES, we applied millimolar range concentrations in PBS for a very short time. Then, cells were incubated 6h in fresh culture medium. We studied the toxicity of the treatment by a MTT assay (Fig. 3). The survival at 24 h was approximately 81.5% ± 4.6 for 5 mM. A much larger lethality was observed at 10 mM with 4.3 % ± 1.0 of viability at 24 h.

< Figure 3 >

*CEES adducts in DNA*

During the treatment, CEES enters the cells where it alkylates many biomolecules including DNA. Different DNA adducts can be formed because CEES reacts on the N7-positions of guanine and the N3-position of adenine. We only quantified N7Gua-CEES, which is the most frequent in CEES-treated cells [47]. Another reason for this choice was the fact that the N7Gua adduct is also the most frequent for SM. This was first indirectly shown by Brookes and Lawley in cell-free systems [53]. More recently, *Batal et al* determined the kinetics of formation of SM-adducts in skin from mice exposed to SM. The most abundant and persistent adduct was also the HETE-N7Gua, which was produced in a dose dependent manner [27]. In addition, Zubel at al investigated the DNA adducts stability in HaCat and A549 cells; N7Gua was still detectable 6 days after treatments with SM [54]. In our work on HaCat cells, the corresponding N7Gua-CEES was detected in DNA under all the investigated conditions, even at the lowest CEES concentration. Formation of N7Gua-CEES adducts was dose-dependent with a linear increase of the quantity of N7Gua-CEES with respect to the CEES concentration (Fig. 4).

< Figure 4 >

*Biomarkers in the medium of HaCaT cells exposed to CEES*

The medium collected from the same experiments was analyzed by UHPLC-MS/MS after SPE clean up. The Cys-CEES and GSH-CEES biomarkers were detectable and quantifiable (Electronic Supplementary Material, Fig. S4). In contrast, NAC-CEES was not detected in this *in vitro* study. The amounts of GSH-CEES and Cys-CEES detected in the medium were dose-dependent (Fig. 5a). At all-time points, the concentration in Cys-CEES was larger than that of GSH-CEES. Interestingly, the ratio between the concentrations of Cys-CEES and GSH-CEES varied from one experimental condition to the other. The value is 4.5 at the lowest CEES concentration and decreases to 2 at the largest one. This is reflected in the dose-concentration curves that is roughly linear for Cys-CEES and more quadratic for GSH-CEES. This trend most likely reflects a saturation of the enzymes involved in the metabolism of GSH conjugates when the CEES concentration increased.

N7Gua-CEES was also detectable in the medium, although in approximately 20 times lower amounts than Cys-CEES and GSH-CEES (Fig. 5b). The presence of this nucleic acid derivative can be explained by different mechanisms. First, It is well known that such N7 DNA adducts are chemically unstable and spontaneously undergo depurination, namely the cleavage of the *N*-glycosidic bond [55]. This process leads to the release of the free alkylated base N7Gua-CEES and the formation of an abasic site in DNA [55, 56]. A similar process takes place in CEES-treated RNA [47] and mostly likely in dGTP from the nucleotide pool. Other sources of DNA adducts in medium could involve DNA repair. The base excision repair pathway leads the release of damaged bases by cleavage of the *N*-glycosidic linkage. Bulky adducts like N7Gua-CEES are also handled by the nucleotide excision repair system which removes a single-stranded DNA fragment bearing the damage [57], from which N7Gua adducts may depurinate.

< Electronic Supplementary Material, Figure S4 >

< Figure 5 >

**Biomarker of exposure to CEES in human skin explants**

The HaCat cell experiments showed that the biomarkers were actually excreted from cultured cells. In order to better understand the response of cutaneous cells to vesicants, we then used whole skin explants that contained both the epidermis and the dermis. We investigated whether the chosen biomarkers where detected after exposure to different amounts of CEES.

< Table 1 >

*DNA adducts from skin exposed to CEES*

Like for HaCat cells, we quantified N7Gua-CEES in nuclear DNA and all biomarkers in the culture medium. Even the lowest concentration applied induced quantifiable amounts of N7Gua-CEES within DNA after 1h of incubation (Table 1). It is well documented that, because of its hydrophobic nature, SM easily penetrates within skin and accumulates in the lipid component of exposed tissues [58]. By analogy, it is expected that in our experiments CEES rapidly penetrates the explants. Since our experiments involved isolated skin without blood flow, the diffusion of CEES out of the tissue is limited and the concentration in CEES is likely to be rapidly stable. This behavior could explain why the amount of N7Gua-CEES does not linearly increase with increasing applied dose during the 24 h of the experiment (Table 1). As illustrated in Figure 6a for the lowest applied amount (80 nmol), the level of DNA adduct actually decreases after 6h. This decrease is certainly explained by the spontaneous depurination and the different DNA repair mechanisms (BER and NER). Interestingly, N7Gua-CEES is detected in the medium 6h after the treatment, but not after 1 h, and its concentration is larger at 24 h. This observation validates the idea that N7Gua-CEES formed in DNA, as well as in RNA and nucleotides, is excreted from cells. It can then be detected as a modified base in the extracellular medium. This confirms the results obtained with cultured HaCat cells and shows the interest of this adduct as a biomarker in biological fluids, in line with the detection of SM-DNA adducts in urine of exposed patients [17, 31].

< Figure 6 >

*Glutathione conjugates from skin exposed to CEES*

Following topical exposure of skin to CEES, the culture medium placed under the insert was collected after different incubation time (1h, 6h 24h). The samples were cleaned up by SPE and analyzed by UHPLC-MS/MS. Cys-CEES and GSH-CEES were detectable and quantifiable, but not NAC-CEES. A first observation is that, for each post-treatment time, the amounts of GSH-CEES and Cys-CEES seemed to saturate when the dose increased (Table 1). This may reflect either a limitation in the intracellular GSH concentration or a saturation of the glutathione-*S*-transferase activity. Temporal profiles of the evolution of the two former compounds were different (Fig. 6b). The concentration in GSH-CEES was maximal at 1h and then decreased while an opposite trend was observed for Cys-CEES. Following exposure, CEES enters into skin cells and, as seen for DNA adducts, rapidly reacts with the thiol function of glutathione to form the GSH-CEES conjugate [59]. GSH-CEES is rapidly excreted, which explained why we detected it in the medium (21.5 ± 10.3 nM) already after 1h. This early time is the maximum of concentration of GSH-CEES since this conjugate is not persistent. It is detectable at 24h only in a few samples for the largest amount of applied CEES. In contrast to GSH-CEES, we observed that the concentration in Cys-CEES in the medium, which is dose-dependent, rose between 1 and 6 h and then reached a plateau until 24 h. These observations are in agreement with known characteristics of the mercapturate pathway involved in the detoxification of electrophilic species such as CEES or SM [35, 40]. In the mercapturate pathway, enzymatic activities sequentially convert GSH-CEES into CysGly-CEES and Cys-CEES. These two consecutive reactions are catalyzed by the membrane-bound-enzymes γ-glutamyl-transferase (GGT) and dipeptidase or aminopeptidase-M, respectively. Our results strongly suggest than the mercapturate pathway seems effective in the skin during the first hours after exposure. However, Cys-CEES may not be produced only by the mercapturate pathway. At the earliest time, its level (125.7 ± 30.8 nM) is already higher than that of GSH-CEES. The largest concentration of Cys-CEES in the medium is approximately one order of magnitude larger than that of GSH-CEES. These results suggest that a part of Cys-CEES detected comes from other processes or biological mechanisms. A direct reaction between the thiol group of intracellular free cysteine is another possible source.

**CEES exposure biomarkers in plasma**

We then applied our approach to samples from an *in vivo* study. Blood plasma from mice cutaneously exposed to vapors of CEES were analyzed by SPE – UHLPC-MS/MS 1 day after exposure. While no peak was observed on the chromatograms corresponding to untreated animals (Fig. 7a), the four biomarkers were readily detected in plasma from treated mice (Fig. 7b). After correction of the peak intensity by the calibration curve, the decreasing concentration of the four analytes was found to be NAC-CEES > Cys-CEES ≈ Gua-CEES >> GSH-CEES (Fig. 8c). The trend observed *in vitro* of a lower amount of GSH-CEES was even more obvious *in vivo*. Gua-CEES was also found to be an interesting biomarker easily accessible from blood, as previously shown in urine or in nuclear DNA extracted from organs and blood cells [17, 31]. The major difference between plasma and cultured cells and skin explants is that NAC-CEES was the most frequent biomarker. NAC-CEES was detected neither in the HaCat cells, skin explants experiments reported above, nor in previous works with primary cultures of normal human keratinocytes (Batal and Douki, unpublished results). The difference between these cellular cutaneous models and the mice situation regarding the formation of this final product of the mercapturate pathway documented for numerous chemicals [41, 42] is that acetylation of cysteine conjugates is not an ubiquitous activity in the organism. This last step of the mercapturate pathway takes place when Cys-*S*-conjugates enter the renal tubular cells and hepatocytes via various transporters including organic anion transport polypeptides and cystine/cysteine transporters [60–62]. Acetylation of the Cys-S-conjugates is carried out by the N-acetyl-transferase NAT8, expressed almost exclusively in the kidney proximal tubular cells, and to a lower extent in the liver [63]. NAC-CEES conjugates synthesized in the kidney are directly released into urine, whereas those synthesized in the liver are excreted into the bile. Mercapturate conjugates are also produced in very low amounts in other tissues and then enter the bloodstream [40, 64]. These observations explain why Cys-CEES was not converted into NAC-CEES in cultured cutaneous cells and skin explants while NAC-CEES was abundant in plasma. They also may explain why Cys-CEES accumulates in *in vitro* systems without being further metabolized and becomes the most frequent of the four studied biomarkers. In addition, the difference between our *in vitro* and *in vivo* data are likely explained by a rapid penetration of CEES through the cutaneous barrier in mice followed by a diffusion towards internal organs where NAC-CEES formation is more efficiently produced than in skin.

< Figure 7 >

## Conclusion

The present work allowed us to develop a method for the simultaneous quantification of four biomarkers of exposure to CEES, namely the N7 guanine adduct and the conjugates to glutathione, cysteine and N-acetyl-cysteine (mercapturic acid). The assay involves sample preparation by SPE and quantification by UHPLC-MS/MS. Isotopic dilution is used to compensate for possible loss of material during the sample work-up and for matrix effects in the MS/MS detection. The accuracy of the method for culture medium samples was validated using reference statistical procedures. Using this technique, we quantified GSH-CEES, Cys-CEES and N7Gua-CEES in the culture medium of exposed HaCat cells and human skin explants. The same compounds and NAC-CEES were detected in plasma from exposed mice. These products of the reaction of CEES with biomolecules are thus unambiguously formed in cells and further excreted. These observations provide a biological validation to the use of GSH-CEES, Cys-CEES, NAC-CEES and N7Gua-CEES as biomarkers of exposure to CEES. They can be thus targeted in biological fluids.

However, not all four compounds are equally interesting as exposure biomarkers. For example, GSH-CEES was found to be rapidly metabolized and is present only in very low amounts in plasma. Cys-CEES could be an interesting marker but the sensitivity of its detection is lower than that for the other compounds. NAC-CEES is a more attractive GSH metabolite since it is present in the largest concentration in plasma and exhibits remarkable analytical properties, like Gua-CEES. These conclusions will be used for the next step of our work involving the development of biomarkers of exposure to SM that will focus and NAC, Gua and to lesser extent Cys derivatives. Another necessary improvement of our approach for real life applications will be the increase of the throughput of the SPE sample preparation step. Although use of SPE cartridges can be automatized, we will rather develop in the future an approach using "on-line SPE". In this system, the SPE column is part of the LC system placed ahead of the mass spectrometer. The sample preparation is thus easier, less time-consuming, and more reproducible.

## Acknowledgements

This work was supported by the "Agence de l'Innovation de Défense" (French Defence Ministry) and the "NRBC" program of CEA.

## Conflict of Interest

The authors declare that they have no conflict of interest.

FIGURES

**Figure 1:** Reactivity of sulfur mustard with biomolecules

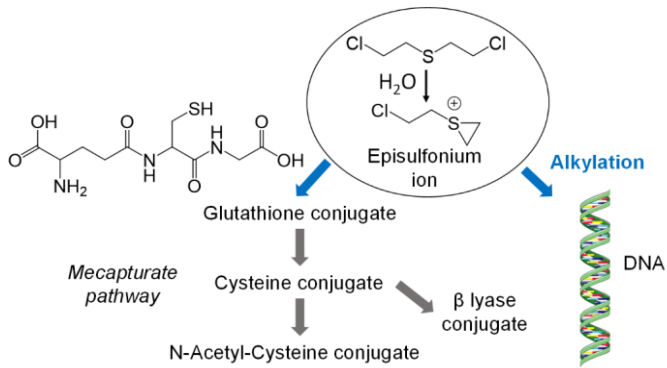

**Figure 2:** The structures of the targeted CEES biomarkers

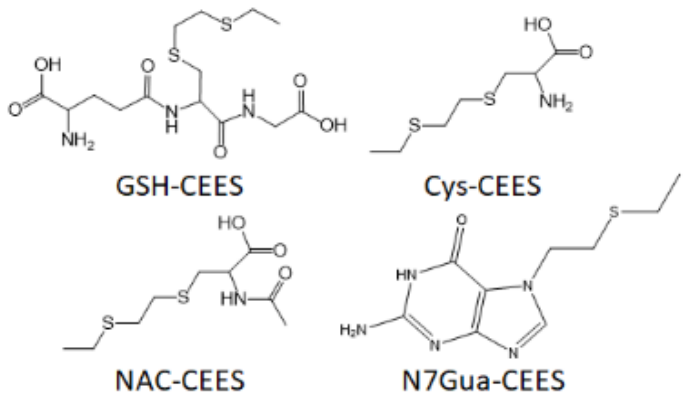

**Figure 3:** Survival of HaCaT cells exposed for 30 min to increasing concentrations of CEES, ranging between 0 and 100 mM. Survival was measured by a MTT assay performed 24 h after treatment. Result are mean ± SEM (n=6)

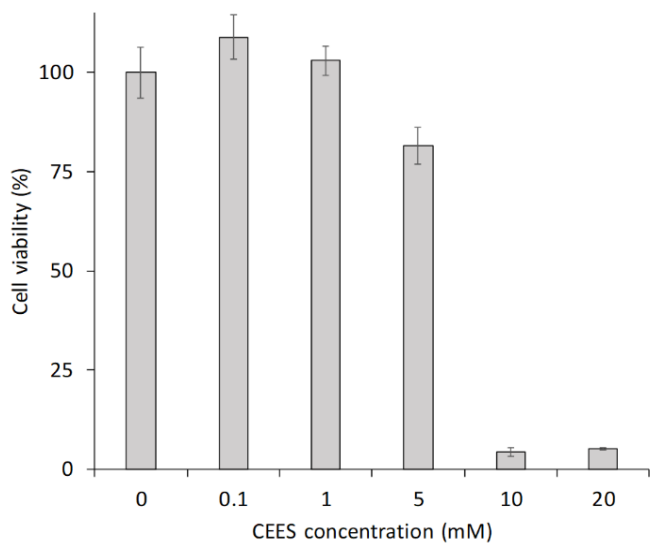

**Figure 4:** Formation of N7Gua-CEES in HacaT cells exposed to CEES concentration ranging between 0 and 10 mM. DNA adducts were quantified by UHPLC-MS/MS (n=9) after DNA extraction and hydrolysis. Results were normalized with respect to the amount of normal bases.

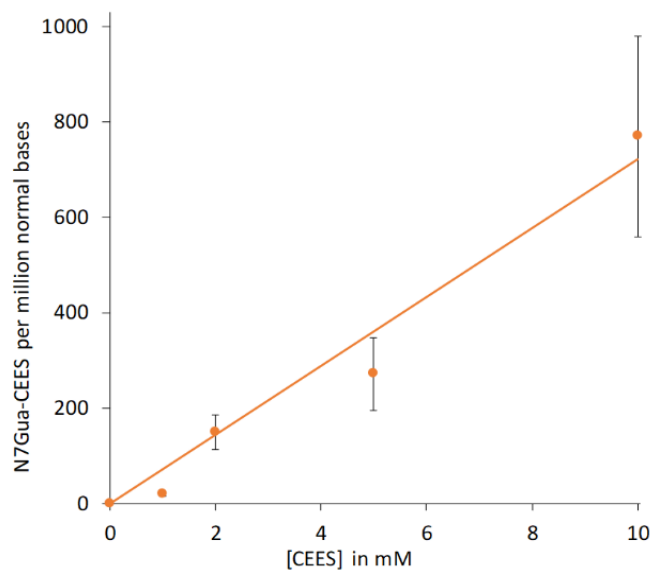

**Figure 5:** Detection and quantification of biomarkers in the culture medium of HaCat cells a) GSH and Cys conjugates, b) N7Gua adduct. Cells were exposed to 30 min to 0, 1, 2, 5, 10 mM of CEES After 6h incubation, the medium was collected, purified by SPE and analyzed by UHPLC-MS/MS (n=9). Results are expressed in concentration of analyte in the culture medium, and represent mean ± standard deviation.

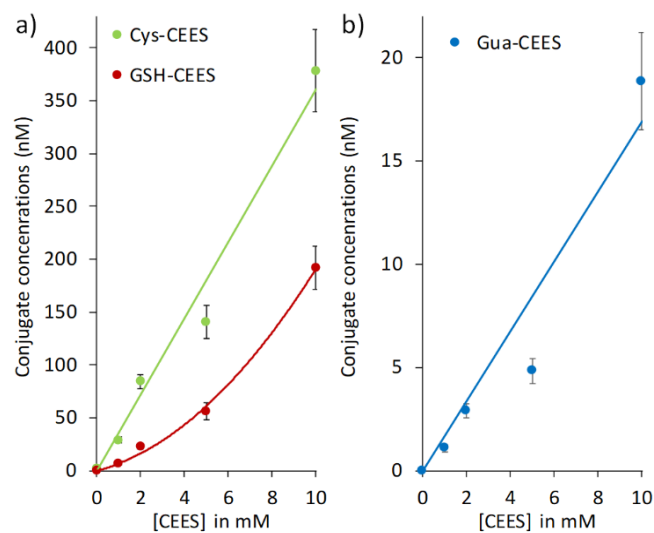

**Figure 6:** Time-course evolution of the level of biomarkers following exposure of human skin to 80 nmol of CEES a) N7Gua-CEES in the DNA and culture medium. DNA adducts were quantified by UHPLC-MS/MS after DNA extraction and digestions (n=6). For DNA, results were normalized with the respect to the amount of normal bases. The media were collected, purified by SPE and analyzed by UHPLC-MS/MS (n=6); b) glutathione conjugates induced by the mercapturate pathway in culture medium. The media were collected, purified by SPE and analyzed by UHPLC-MS/MS (n=6).

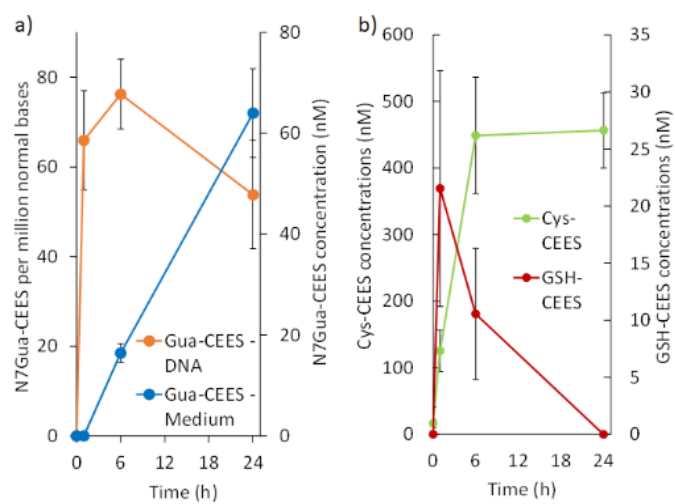

**Figure 7:** Detection of the biomarkers in blood plasma from mice. Chromatogram of the analysis of plasma from a) unexposed mice and b) mice 1 day after cutaneously exposure to vapors of CEES; c) concentration of the four biomarkers in the plasma of exposed mice. Results are expressed in pmol/µl. They are means ± SEM (n=4)

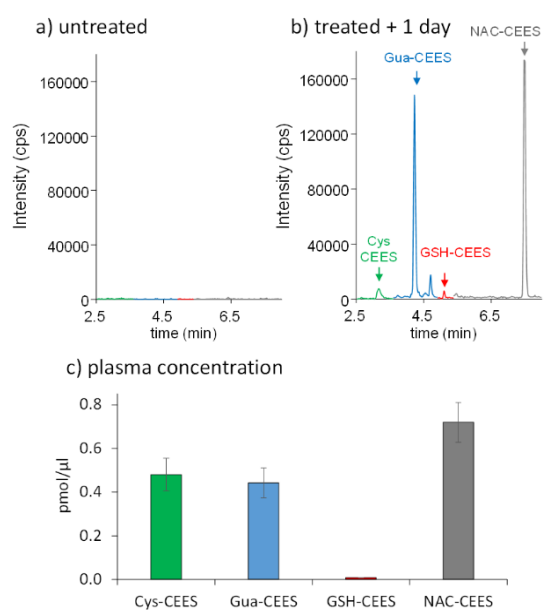

# Glutathione conjugates of the mercapturic acid pathway and guanine adduct as biomarkers of exposure to CEES, a sulfur mustard analog.

By Marie Roser, David Béal, Camille Eldin, Leslie Gudimard, Fanny Caffin, Fanny Gros-Désormeaux, Daniel Léonço, François Fenaille, Christophe Junot, Christophe Piérard and Thierry Douki

## Electronic Supplementary Material

**Table S1:** MRM transitions used for the UHPLC-MS/MS detection of the biomarkers and their internal standards, and the corresponding optimized mass spectrometry parameters.

| Name | Transition (m/z) | DP (V) | EP (V) | CE (V) | CxP (V) | Dwell Time (msec) |
|---|---|---|---|---|---|---|
| Gua-CEES 61 | 240 → 61 | 50 | 8 | 50 | 10 | 50 |
| Gua-CEES 89 | 240 → 89 | 45 | 15 | 20 | 10 | 50 |
| Gua-CEES 55 | 240 → 55 | 45 | 8 | 50 | 5 | 50 |
| *Gua-CEES\* ID 61* | *245 → 61* | 50 | 8 | 50 | 10 | 50 |
| *Gua-CEES\* ID 89* | *245 → 89* | 45 | 15 | 20 | 10 | 50 |
| *Gua-CEES\* ID 55* | *245 → 55* | 45 | 8 | 50 | 5 | 50 |
| GSH-CEES 89 | 396 → 89 | 45 | 8 | 20 | 10 | 50 |
| GSH-CEES 121 | 396 → 121 | 45 | 8 | 20 | 10 | 50 |
| GSH-CEES 321 | 396 → 321 | 45 | 8 | 20 | 10 | 50 |
| *GSH-CEES\* ID 61* | *399 → 61* | 45 | 8 | 70 | 10 | 50 |
| *GSH-CEES\* ID 89* | *399 → 89* | 45 | 8 | 20 | 10 | 50 |
| *GSH-CEES\* ID 121* | *399 → 121* | 45 | 8 | 20 | 10 | 50 |
| Cys-CEES 61 | 210 → 61 | 45 | 8 | 60 | 10 | 50 |
| Cys-CEES 89 | 210 → 89 | 45 | 8 | 20 | 10 | 50 |
| Cys-CEES 120 | 210 → 120 | 45 | 8 | 20 | 5 | 50 |
| *Cys-CEES\* ID 61* | *214 → 61* | 45 | 8 | 60 | 10 | 50 |
| *Cys-CEES\* ID 89* | *214 → 89* | 45 | 8 | 20 | 10 | 50 |
| *Cys-CEES\* ID 93* | *214 → 93* | 45 | 8 | 20 | 5 | 50 |
| NAC-CEES 61 | 252 → 61 | 45 | 8 | 60 | 5 | 50 |
| NAC-CEES 89 | 252 → 89 | 45 | 8 | 10 | 5 | 50 |
| NAC-CEES 116 | 252 → 116 | 45 | 15 | 50 | 5 | 50 |
| *NAC-CEES\* ID 61* | *256 → 61* | 45 | 8 | 60 | 5 | 50 |
| *NAC-CEES\* ID 89* | *256 → 89* | 45 | 8 | 10 | 5 | 50 |
| *NAC-CEES\* ID 119* | *256 → 119* | 45 | 15 | 50 | 5 | 50 |

**Table S2:** [1]H-NMR characterization of the GSH, NAC and Cys conjugates to CEES. Mobile protons were first replaced by deuterium by freeze-drying in D$_2$O. Samples were then solubilized in 99.95 % D$_2$O and the spectra recorded at 400 MHz. Identification of the signals was based on the analys1s of the coupling constants (*J*) and the predicted spectra generated by the ChemDraw software. Additional information were obtained for GSH-CEES by comparison with published data for the corresponding the sulfur mustard conjugate [46].

| Cys-CEES | | 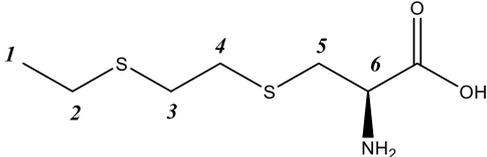 | | |
|---|---|---|---|---|
| Chemical shift (ppm) | Integration | Multiplicity | *J* (Hz) | Proton |
| 1.17 | 3H | t | 7.4 | 1 |
| 2.56 | 2H | q | 7.4 | 2 |
| 2.79 | 4H | t | 3.1 | 3 & 4 |
| 3.05 | 2H | m | / | 5 |
| 3.87 | 1H | dd | 4.4 / 7.3 | 6 |
| NAC-CEES | | 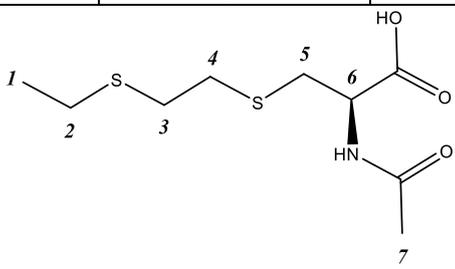 | | |
| Chemical shift (ppm) | Integration | Multiplicity | *J* (Hz) | Proton |
| 1.17 | 3H | t | 7.4 | *1* |
| 2.00 | 3H | s | / | *7* |
| 2.55 | 2H | q | 7.2 | *2* |
| 2.76 | 4H | m | / | *3 & 4* |
| 3.02 & 2.87 | 2H | m | / | *5* |
| 4.36 | 1H | dd | 4.6 / 8 | *6* |
| GSH-CEES | | 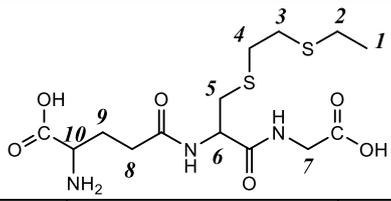 | | |
| Chemical shift (ppm) | Integration | Multiplicity | *J* (Hz) | Proton |
| 1.17 | 3H | m | / | *1* |
| 2.11 | 2H | m | / | *9* |
| 2.55 | 4H | m | / | *2 & 8* |
| 2.76 | 4H | m | / | *3 & 4* |
| 2.85 | 1H | dd | 9 / 14 | 5 |
| 3.05 | 1H | dd | 5 / 14 | 5 |
| 3.7 | 2H | m | / | 7 |
| 4.01 | 1H | m | / | 10 |
| 4.53 | 1H | dd | 4.6 / 9.1 | 6 |

**Table S3:** Statistical validation of the method. Calculated concentration, accuracy and precision are reported for calibrators at the lowest and highest concentration. Conc.: concentration, Calc. Conc.: Calculated concentration.

|  | Sample | Conc. (µM) | Intra-day (n=6) | | | Inter-day (n=18) | | |
|---|---|---|---|---|---|---|---|---|
|  |  |  | Calc. Conc. (µM) | Accuracy (%) | RSD (%) | Calc. Conc. (µM) | Accuracy (%) | RSD (%) |
| GSH-CEES | Calibrator. 1 | 0.1 | 0.11 | 111 | 9.6 | 0.10 | 101 | 13.9 |
|  | Calibrator 6 | 50 | 49.3 | 99 | 4.0 | 496 | 99 | 3.9 |
| Cys-CEES | Calibrator 1 | 0.1 | 0.11 | 113 | 5.5 | 0.11 | 105 | 7.5 |
|  | Calibrator 6 | 50 | 49.2 | 98 | 4.7 | 49.8 | 100 | 4.6 |
| NAC-CEES | Calibrator 1 | 0.1 | 0.11 | 114 | 8.2 | 0.10 | 104 | 11.8 |
|  | Calibrator 6 | 50 | 49.5 | 99 | 5.8 | 49.8 | 100 | 5.5 |
| Gua-CEES | Calibrator 1 | 0.1 | 0.10 | 98 | 9.8 | 0.09 | 87.3 | 17.5 |
|  | Calibrator 3 | 0.5 | 0.51 | 101 | 7.6 | 0.50 | 99 | 7.1 |

**Table S4:** Intra-day validation of the quantification of the biomarkers in mouse plasma. Aliquot fractions of samples from mice exposed to CEES (day 1) were mixed to obtain an overall volume of 500 µL. The resulting solution was spiked with GSH-CEES the concentration of which is very low in plasma. The final concentration of the analytes was approximately 50 nM. Four aliquot fractions of the mixture were then purified by SPE and analyzed by UHPLC-MS/MS as described in the manuscript.

| analyte | GSH-CEES | Cys-CEES | NAC-CEES | Gua-CEES |
|---|---|---|---|---|
| RSD (%) | 5.7 | 9.2 | 12.4 | 9.0 |

**Figure S1:** Protocol of CEES exposure on hairless SKH-1 mice using vapor caps method. Caps containing CEES were maintained 30 minutes on back skin. For quantification of biomarkers, mice were sacrificed 1 day after CEES vapor exposure. Macroscopic investigation of the burn was followed up to 7 days. The wounds reached their maximal intensity at day 3 (D3).

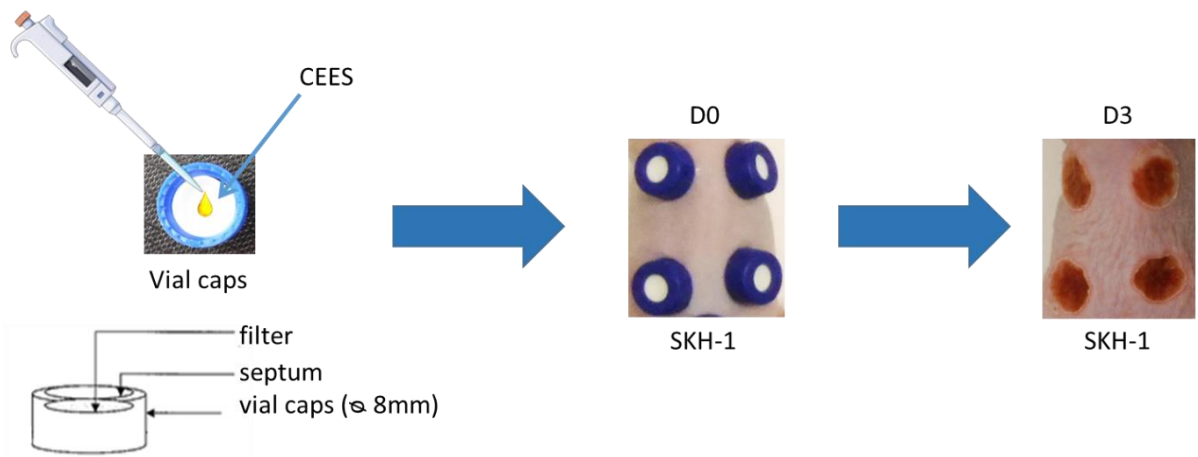

**Figure S2:** Fragmentation mass spectra of the targeted analytes and identification of the main fragments used for the UHPLC-MS/MS quantification.

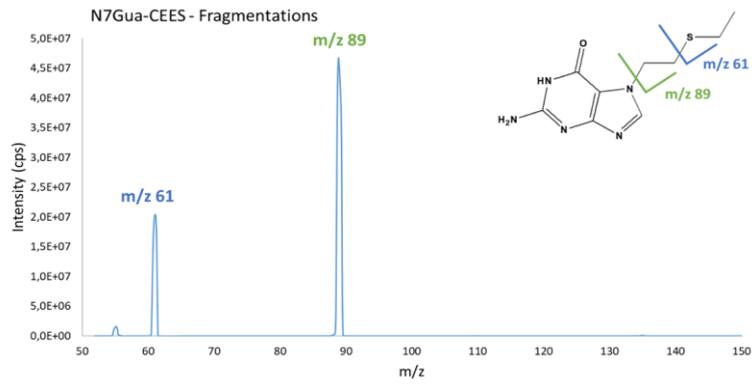
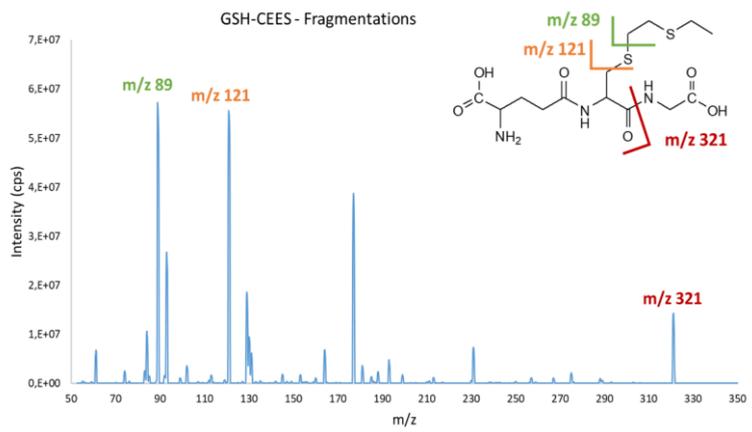
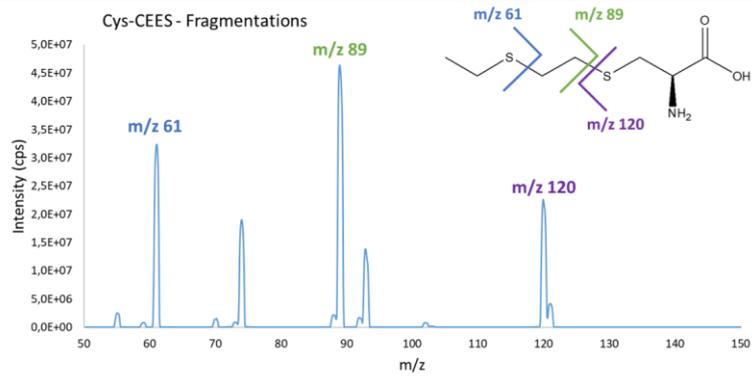
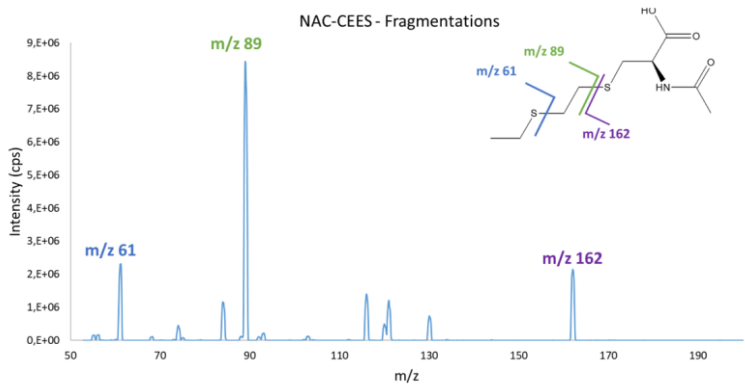

**Figure S3:** UHPLC chromatograms recorded in the product ion scan mode corresponding to the detection of Cys-CEES (210 m/z), Cys-CEES sulfoxide (226 m/z), Cys-CEES sulfone (242 m/z) in culture medium from skin exposed to 0.40 µmol of CEES during 0h, 1h, 6h or 24h. No peak at M+16 or M+32 with respect to Cys-CEES is observed.

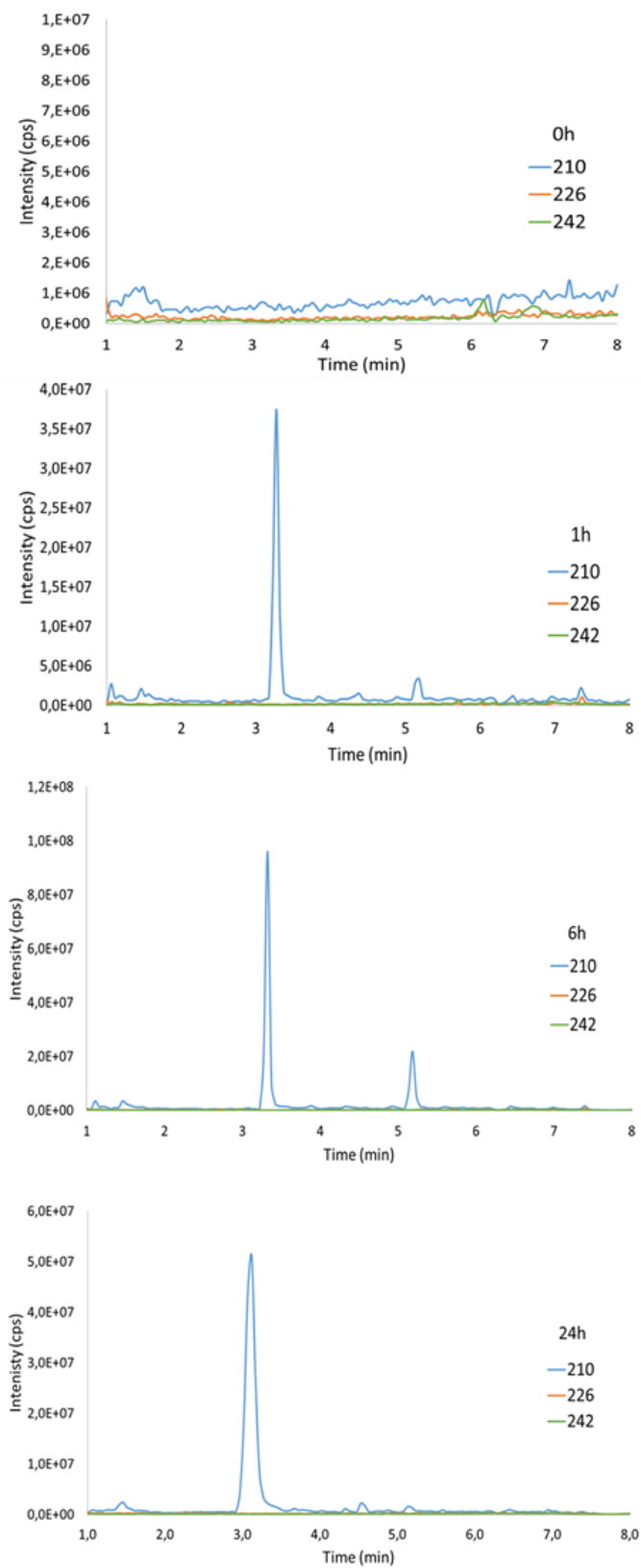

**Figure S4:** UHPLC-MS/MS MRM chromatogram of the analysis of N7Gua-CEES, Cys-CEES, GSH-CEES and NAC-CEES biomarkers determined in culture medium from HaCaT exposed to 1mM of CEES during 30min.

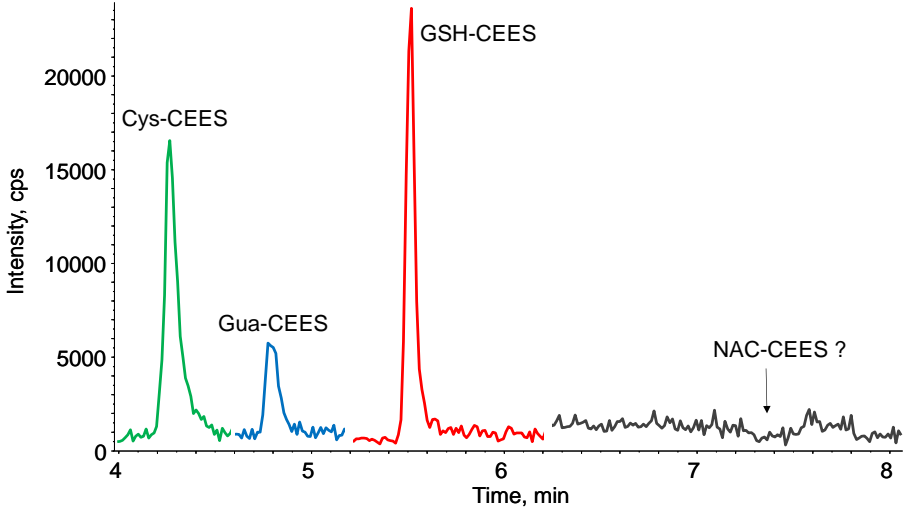